\documentclass[
    ,final            
  ]
  {aipproc}

\layoutstyle{8x11single}

\newcommand{\apj}{ApJ}

\newcommand{\aap}{A\&A}
\newcommand{\mnras}{MNRAS}
\newcommand{\nat}{Nature}

\begin{document}

\title{\emph{Swift} GRBs and the blast wave model}





\author{P.A.~Curran\thanks{e-mail: pac@mssl.ucl.ac.uk}  }{
  address={Mullard Space Science Laboratory, University College London, Holmbury St. Mary, Dorking RH5~6NT, UK}
  ,altaddress={Astronomical Institute, University of Amsterdam, Kruislaan 403, 1098\,SJ Amsterdam, The Netherlands}
}

\author{A.J.~van~der~Horst}{
  address={NASA Postdoctoral Program Fellow, NSSTC, 320 Sparkman Drive, Huntsville, AL 35805, USA}
}

\author{R.L.C.~Starling}{
  address={Department~of~Physics~and~Astronomy, University~of~Leicester, University~Road, Leicester~LE1~7RH, UK}
}

\author{R.A.M.J.~Wijers}{
  address={Astronomical Institute, University of Amsterdam, Kruislaan 403, 1098\,SJ Amsterdam, The Netherlands}
}

\begin{abstract}

The complex structure of the light curves of \emph{Swift} GRBs has made their interpretation and that of the blast wave caused by the burst, more difficult than in the pre-\emph{Swift} era. We aim to constrain the blast wave parameters: electron energy distribution, $p$, density profile of the circumburst medium, $k$, and the continued energy injection index, $q$. We do so by comparing the observed multi-wavelength light curves and X-ray spectra of a \emph{Swift} sample to the predictions of the blast wave model.

We can successfully interpret all of the bursts in our sample of 10, except two, within the framework of the blast wave model, and we can estimate with confidence the electron energy distribution index for 6 of the sample. Furthermore  we identify jet breaks in half of the bursts.
A statistical analysis of the distribution of $p$ reveals that, even in the most conservative case of least scatter, the values are not consistent with a single, universal value.  The values of $k$ suggest that the circumburst density profiles are not drawn from only one of the constant density or wind-like media populations. 

\end{abstract}

\maketitle


\section{Introduction}\label{introduction}

The afterglow emission of Gamma-Ray Bursts (GRBs) is generally well described by the blast wave model \citep{rees1992:MNRAS258,meszaros1998:ApJ499}. This model details the temporal and spectral behaviour of the emission that is created by external shocks when a collimated ultra-relativistic jet ploughs into the circumburst medium, driving a blast wave ahead of it. 
The level of collimation, or jet opening angle, has important implications for the energetics of the underlying physical process, progenitor models, and the possible use of GRBs as standard candles. The signature of this collimation is an achromatic temporal steepening  or `jet break' at approximately one day in an otherwise decaying, power-law light curve.

Since the launch of the \emph{Swift} satellite it has become clear that this model for GRBs cannot, in its current form, explain the full complexity of observed light curve features and the lack of observed achromatic temporal breaks. The unexpected features detected, such as steep decays, plateau phases (e.g., \cite{tagliaferri2005:Natur436,nousek2006:ApJ642,obrien2006:ApJ647}) and a large number of X-ray flares (e.g., \cite{burrows2007:RSPT365,chincarini2007:ApJ671}) have revealed the complexity of these sources up to about one day since the initial event, which is yet to be fully understood. These superimposed features also make it difficult to measure the underlying power-law features on which the blast wave model is based, and may lead to misinterpretations of the afterglows.

In these proceedings we summarize our interpretation of a sample of 10 \emph{Swift} GRB afterglows which we detail in our paper \cite{curran2009:MNRAS}. Here, we introduce our method of sample selection and analysis, and summarize our main results regarding the constraints we can place on the blast wave parameters: electron energy distribution, $p$, density profile of the circumburst medium, $k$, and the continued energy injection index, $q$. 
Throughout, we use the convention that a power-law flux is given as $F \propto t^{-\alpha} \nu^{-\beta}$ where $\alpha$ is the temporal decay index and $\beta$ is the spectral index.  


\section{Sample and analyses}\label{observations}

The bursts in our sample were chosen from an inspection of previous literature and from a comparison of the literature of optical data to the pre-reduced \emph{Swift} X-ray Telescope (XRT) light curves in the on-line repository \cite{evans2007:A&A469} up to the end of February  2008. Our sample consists of 10 bursts with X-ray and optical light curves with good enough time coverage to allow for the underlying single power-law, or broken power-law, to be determined. The bursts are also well sampled enough in the X-ray to constrain the spectral indices, $\beta_{{\rm X}}$.  We did not confine our sample to bursts with clear breaks in either the X-ray or optical bands as we wanted to include the possibility of hidden or not very obvious breaks, particularly in the X-ray band \citep{curran2008:MNRAS386}, or late, undetected breaks.

\begin{figure}[h] \label{lc}
 \includegraphics[height=.5\textheight,angle=270]{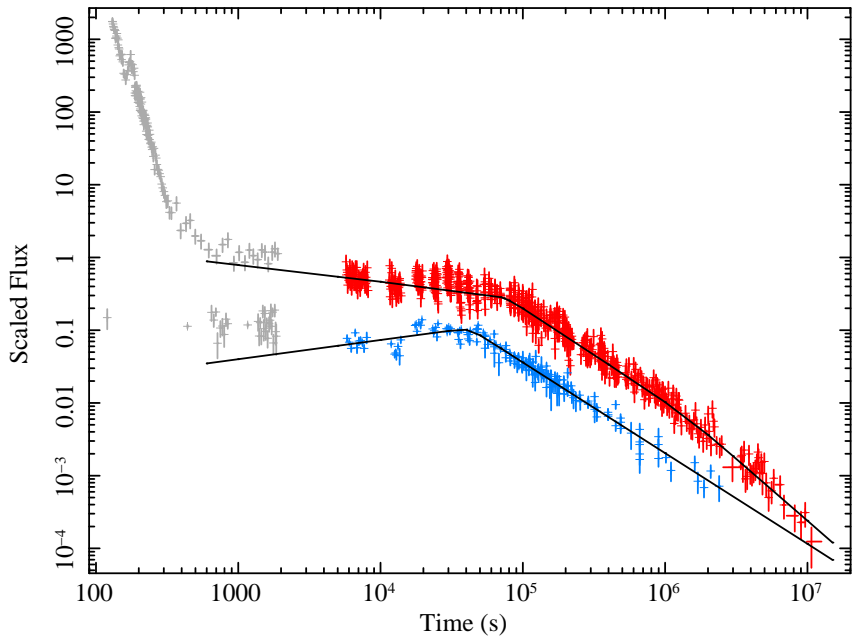}
\caption{The power-law fits to the XRT (upper data) and optical (lower data) light curves of an example burst from our sample (GRB\,060729). Early time, flaring data were not used in the power-law fits.}
\end{figure}

Light curve analyses were carried out on the pre-reduced, XRT light curves from the on-line repository. For bursts where there was a possible light curve break, X-ray spectra were extracted pre-break and post-break.  Optical photometric points in various bands were taken from the literature and combined via a simultaneous temporal fit. This fitting allowed us to find the common temporal slope of the optical data and the colour differences between bands.  Using these colours, the optical data were then shifted to a common magnitude and converted into an arbitrary, scaled flux to produce joint optical and X-ray light curves (Figure\,\ref{lc}). These light curves were fit with single or broken power-laws, including optical host galaxy contributions where known. Data at early times at which the underlying behaviour was ambiguous, or flaring, were excluded from the fit.

%
%
%

\subsection{Derivation of electron energy distribution index, $p$}\label{p-derivation}

We use the blast wave model \cite{rees1992:MNRAS258,meszaros1998:ApJ499} to describe the temporal and spectral properties of the GRB afterglow emission; we assume on-axis viewing, a uniform jet structure and no evolution of the microphysical parameters. The relations between the temporal and spectral indices and the blast wave parameters that we use are summarised in, e.g., \cite{nousek2006:ApJ642,starling2008:ApJ672}. 
Our general method was to estimate the value of the electron energy distribution index, $p$,  from the X-ray spectral index and use this to calculate the predicted values of temporal decay. 
We derive $p$ from the spectral index as opposed to the temporal index since for a given spectral index there are only two possible values of $p$, while for a given temporal index there are multiple possible values. Spectral slopes are dependent only on $p$ and the position of the cooling break. Temporal indices, $\alpha$,  are dependent on  $p$, the position of the cooling break, the circumburst density profile, $k$, and on possible continued energy injection. Temporal indices are also prone to being incorrectly estimated from broken power-law fits which may underestimate the post-break indices \citep{johannesson2006:ApJ640}.

For a given value of the X-ray spectral index, there are two possible values of $p$ depending on whether the cooling break, $\nu_{{\rm c}}$, is below ($p = 2\beta$) or above ($p = 2\beta +1$) the X-ray frequency, $\nu_{{\rm X}}$. If the optical to X-ray SED  does not display a break then the cooling break can either be above the X-ray regime or below the optical regime and the blast wave predictions of each $p$ are compared to the observed temporal slopes to discern which is correct. If the SED requires a broken power-law it most likely implies that a cooling break lies between the two regimes and is below the X-ray regime. A cooling break requires, or must be consistent with, a difference between the spectral slopes of $\Delta\beta = 0.5$. 
However, a break between the two regimes does not necessarily imply a cooling break; it may be due to the fact that each regime has a different spectral index since they are originating from different emission regions. In this case the spectral break does not have a predictable difference between slopes. 
For this interpretation to work, one must be able to explain why the emission from each region is only visible in one spectral regime and it's power-law slope does not extend to the other. A cooling break is a more likely explanation in the majority of cases but a comparison of the blast wave predictions of each $p$ with the observed light curves is required.

\section{Results}\label{results}

After comparing the derived values of $p$ and the predictions of the blast wave model to the observed temporal and spectral properties, we find the most likely values of the parameters $p$, $k$ and $q$ for each burst in our sample. We find that: 

\renewcommand{\labelitemi}{$\star$}
\begin{itemize}
\item{8/10 are consistent with the blast wave model}
\item{6/10 have an unambiguous value of $p$}
\item{6/10 have a calculable value of $k$}
\item{4/10 require energy injection, $q$}
\item{5/10 exhibit a jet break}
\end{itemize}

\subsection{Distribution of $p$}\label{p-distribution}

\begin{figure}[b] \label{p}
 \includegraphics[height=.4\textheight,angle=270]{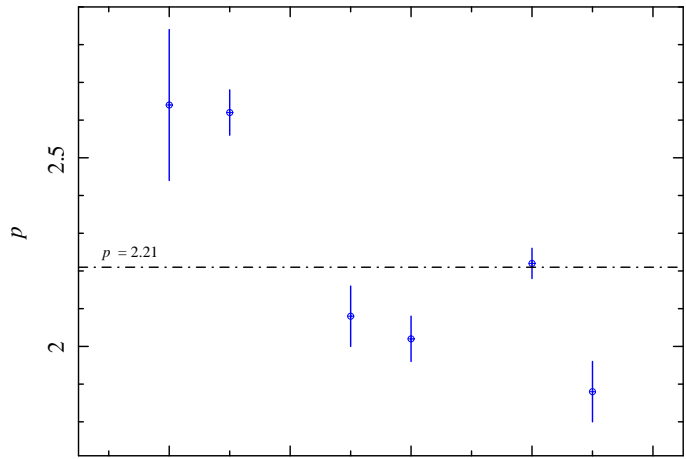}
\caption{Most likely values for $p$ after our interpretation of individual bursts. The line represents the most likely value of $p$ over the plotted sample.}
\end{figure}

The universality of the electron energy distribution index, $p$, has been examined by several authors \cite{chevalier2000:ApJ536,panaitescu2002:ApJ571,shen2006:MNRAS371,starling2008:ApJ672} who applied different methods to samples of \emph{BeppoSAX} and \emph{Swift} bursts, all reaching the conclusion that the observed range of $p$ values is not consistent with a single central value of $p$ but displays a width of the parent distribution. In the studies so far there have been some limitations: the \emph{BeppoSAX} sample is limited, both in the number of GRBs and the temporal and spectral sampling; and the only study of \emph{Swift} bursts for this purpose so far \cite{shen2006:MNRAS371}, only used the X-ray afterglows, which introduces a large uncertainty because the position of the cooling break is unknown.

Here we examine the universality of $p$ given the observed distribution of $p$ (Figure\,\ref{p}) from our sample of \emph{Swift} bursts, using the same methods as described in \cite{starling2008:ApJ672}. We first find the most likely value of $p=2.21\pm 0.03$ and test that the observed distribution can be obtained from a parent distribution with a single central value of $p$. We do so by generating different synthetic sets of $p$ for the bursts in our sample and calculating the most likely value of $p$ for each. We conclude that a single value of $p$ is rejected at the $>5\,\sigma$ level and that the width of the parent distribution is $\sigma_{\rm{scat}} \sim 0.3$ at the $1\,\sigma$ level.

We also tested the values of $p$, for each of the 10 bursts in our sample, that offered the least deviation from the expected canonical value of $\sim 2.2$. In this case, the most likely value of $p$ is $2.19 \pm 0.02$ and the values are still inconsistent with one central value at the $5\sigma$ level.

This result confirms the results from previous studies and has important implications for theoretical particle acceleration studies. Some of these (semi-)analytical calculations and simulations indicate that there is a nearly universal value of $p \sim 2.2-2.3$ (e.g. \cite{kirk2000:ApJ542,achterberg2001:MNRAS328}), while other studies suggest that there is a large range of possible values for $p$ of $1.5-4$ \cite{baring2004:NuPhS136}. 
Although we find that there is not a universal value of $p$, our values for the width of the parent distribution indicate that it is not as wide as the latter study suggest. Our result is comparable to the numbers found by other authors \cite{shen2006:MNRAS371,starling2008:ApJ672} but is based on a sample with better temporal and spectral sampling per GRB, on average.

\subsection{Circumburst density profile, $k$}\label{medium}

The density structure, or profile, of the circumburst medium is generally given as $\rho$, or $n$ (number density), $\propto r^{-k}$ where $k=0$ is a constant density, or ISM-like, medium and $k=2$ is a wind-like medium. The value of $k$ has important implications for the study of progenitor models, as the currently favoured model, involving the collapse of a massive, Wolf-Rayet star, is expected to have an associated strong stellar wind affecting the circumburst environment. However, detailed broadband modeling studies on a small number of GRBs  \cite{panaitescu2002:ApJ571,starling2008:ApJ672} have found that although such a wind is favoured in many cases, a constant density medium is favoured in many other cases.

\begin{figure}[h] \label{k}
 \includegraphics[height=.4\textheight,angle=270]{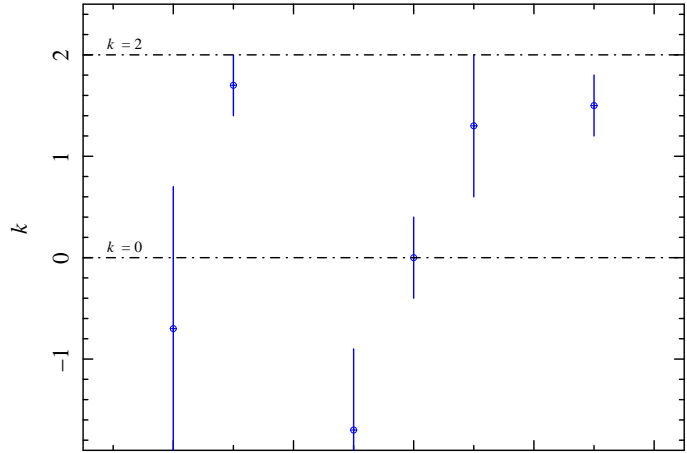}
\caption{ Calculated values of $k$ after our interpretation of individual bursts. The lines represent $k=0$ (constant density) and $k=2$ (wind like) media.}
\end{figure}

In our sample of 10 bursts, only 6 have optical or X-ray light curves below the cooling break, $\nu_{{\rm c}}$, where they are dependent on the circumburst density profile. Of our calculated values of $k$ (Figure\,\ref{k}), 2 are consistent ($2\sigma$) with both $k=2$ and $k=0$; 2 are best described with a wind-like medium and are inconsistent ($5\sigma$) with $k=0$; and 2 are consistent with a constant density medium but inconsistent with $k=2$.  Our results are hence in agreement with those of the previous, similar studies insofar as the sample requires both constant density and wind driven media to explain the observed broadband emission.

\subsection{Rate of continued energy injection, $q$}\label{injection}

\begin{figure}[!] \label{q}
 \includegraphics[height=.4\textheight,angle=270]{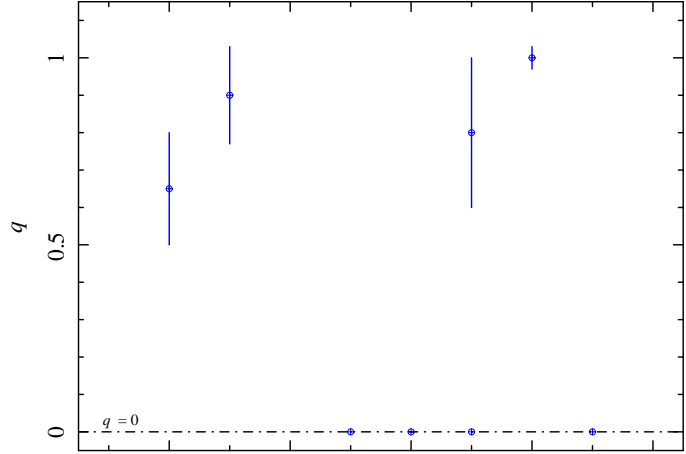}
\caption{Calculated values of $q$, including those with $q$ set as zero (no energy injection) on line, after our interpretation of individual bursts.}
\end{figure}

Continued energy injection was proposed as a process to explain the shallower than expected decay indices observed in many \emph{Swift} bursts \cite{nousek2006:ApJ642}. We assume that it takes the form  of $E \propto t^{q}$ where $q$ is dependent on the shallowing effect of the injection, $\Delta\alpha = \alpha_{p} - \alpha_{{\rm observed}}$, and on the value of $p$. This continued energy injection is either due to $i)$ a distribution of the Lorentz factors of the shells ejected from the central engine which causes those shells with lower Lorentz factor to catch up with the blast wave at a later time, or $ii)$ continued activity of the central engine itself. The former has a limit of $q \leq 3-k$, where $k$ is the density profile of the circumburst medium, which has been suggested as a diagnostic to differentiate between the two sources of continued energy injection \cite{nousek2006:ApJ642}.

For 6 of the 10 light curves in our sample we are able to estimate the level of energy injection, or to say that it is consistent with zero (Figure\,\ref{q}), though in one burst there are two equally valid interpretations, one requiring energy injection while the other does not. Excluding this burst, energy injection is only required in 3 out of 6 bursts and each of these has an injection index, $q$, of between 0.65 and 1.0, consistent with both scenarios of energy injection. Clearly a much larger sample of afterglows is required to be able to constrain the sample properties of $q$.

\section{Conclusions}\label{conclusion}

Throughout this paper we have applied the blast wave model \citep{rees1992:MNRAS258,meszaros1998:ApJ499}, assuming on-axis viewing, a standard jet structure and no evolution of the microphysical parameters, to a selection of 10 \emph{Swift} GRBs with well sampled temporal and spectral data. We attempted to constrain three parameters of interest ($p$, $k$ and $q$) and to test the validity of the blast wave model for light curves observed by \emph{Swift}, the complexity of which has made their interpretation and that of the blast wave more difficult than in the pre-\emph{Swift} era. We find that the majority of the afterglows are well described within the frame work of the blast wave model and that the parameters derived are consistent with those values found by previous authors. Furthermore, we identify, reasonably unambiguously, jet breaks in 5 afterglows out of our sample of 10. 

After interpretation within the blast wave model,  we are able to confidently estimate the electron energy distribution index, $p$, for 6 of the bursts in our sample. A statistical analysis of the distribution of $p$ reveals that, even in the most conservative case of least scatter, the values are not consistent with a single, universal value suggested by some studies; this has important implications for theoretical particle acceleration studies. 
In a number of cases, we are also able to obtain values for the circumburst density profile index, $k$, and the index of continued energy injection, $q$. The calculated values of $q$ are consistent with both suggested sources of continued energy injection and a much larger sample of afterglows will be required to constrain the sample properties of $q$. The values of $k$, consistent with previous works on the matter, suggest that the circumburst density profiles are not drawn from only one of the constant density or wind-like media populations.


\begin{theacknowledgments}
We thank P.A. Evans \& K.L. Page for useful discussions on the XRT. 
PAC \& RAMJW gratefully acknowledge support of NWO under Vici grant 639.043.302.
PAC \& RLCS  acknowledge support from STFC.
AJvdH was supported by an appointment to the NASA Postdoctoral Program at the MSFC, 
administered by Oak Ridge Associated Universities through a contract with NASA.
\end{theacknowledgments}


\end{document}